\documentclass[apl,aip,reprint]{revtex4-1}
\usepackage{amsmath,amssymb}
\usepackage{graphicx}
\usepackage{dcolumn}
\usepackage{bm}
\usepackage{hyperref} 

\begin{document}

\title{Optical transitions and energy relaxation of hot carriers in Si nanocrystals}
\author{A.~N.~Poddubny, A.~A.~Prokofiev, and I.~N.~Yassievich}
\affiliation{Ioffe Physical-Technical Institute RAS, 
26 Polytechnicheskaya, 194021 Saint-Petersburg, Russia}
\date{\today}

\begin{abstract}
Dynamics of hot carriers confined in Si nanocrystals is studied theoretically using atomistic tight binding approach. Radiative, Auger-like and phonon-assisted processes are considered.
The Auger-like energy exchange between electrons and holes is found to be the fastest process in the system.  
However the energy relaxation of hot electron-hole pair is governed by the single optical phonon emission. 
For a considerable number of states in small nanocrystals single-phonon processes are ruled out by energy conservation law.
\end{abstract}

\maketitle

Silicon nanocrystals (NCs) produced by various techniques 
have been of great interest during the past decade.\cite{Pavesi,Kenyon,Katya} 
Observations of effective high energy photoluminescence emission, 
tunable by NC size,\cite{JETF} 
and photon cutting in Si NCs embedded into SiO$_{2}$ matrix \cite{Nature} 
have risen the problem of competition between optical and nonradiative transitions in Si NCs. 
Here we present a theoretical study of radiative recombination and nonradiative relaxation 
of hot confined electrons and holes.

To calculate the states of confined carriers  we use the
$sp^{3}d^{5}s^\ast$ empirical tight binding technique
with nearest neighbor interactions from Ref.~\onlinecite{Jancu}.
Giving the band structure of bulk silicon with the same precision as 
the third nearest neighbor approach does,\cite{Delerue_2000}
this technique, being restricted to the nearest neighbors, 
is more suitable to analyze electron-phonon interaction.
The spin-orbit interaction is neglected in our consideration.

\begin{figure}[b!]
\begin{center}
\includegraphics[width=0.4\textwidth]{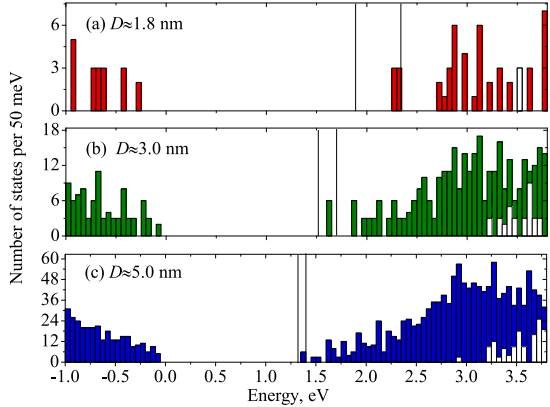}
\end{center}
\caption{Density of states (number of levels per $50$~meV) 
calculated for hydrogenated Si NCs with diameters  close to $1.8$~nm (a), $3.0$~nm (b), $5.0$~nm (c).	
Energies of first two electron states calculated in effective mass approximation,\cite{Moskalenko_PRB2007} including effect of tunneling  into SiO$_{2}$ matrix, 
are shown by vertical dashed lines.
Empty bars correspond to the $\Gamma$-related states.\cite{GammaRelatedStates}}
\label{fig:1}
\end{figure}

We have produced calculations for silicon clusters passivated by hydrogen.
Such boundary conditions are acceptable in the tight binding model~\cite{Delerue_book_2004,Trani2005}
and correspond to the free-standing Si NCs, realized recently.\cite{Kovalev2010}
The parameters for the Si-H bond are taken from Ref.~\onlinecite{Hill}. 
In Fig.~1 we present the calculated density of states for the clusters 
Si$_{147}$H$_{148}$,  Si$_{729}$H$_{444}$ and Si$_{2563}$H$_{1068}$, 
with diameters $D$ being approximately equal to 1.8, 3 and 5 nm, respectively. 
The model clusters are of $T_d$ symmetry and are built by adding atomic layers 
one by one, around the central atom. 

The tight binding method goes beyond effective mass approach,
taking into account the valley-orbit splitting, 
which is up to 0.1~eV for the ground state of electrons when $D\lesssim 2$~nm. 
However, this state-of-the-art method 
does not describe the effect of the carrier tunneling into SiO$_2$. 
More realistic description of the boundary between Si NCs and SiO$_2$ 
in the framework of the tight binding approach is under development now. 
To estimate the effect of tunneling, 
we also show by vertical lines in Fig.~1 
the first two energy levels for confined electrons 
in spherical Si quantum dots embedded in SiO$_{2}$ matrix
obtained in the framework of multi-band effective mass theory.~\cite{Moskalenko_PRB2007}
In Ref.~\onlinecite{Moskalenko_PRB2007} the boundary conditions 
were the continuities of wave function and flux at the interface. 
Both approaches agree for large NCs where tunneling may be neglected.

The probability of direct (phononless) absorption of photons by NCs embedded into
dielectric matrix is given by
\begin{equation}\label{Wabs}
W_{\rm abs}(\hbar\Omega)=\sum_{e,h}\hbar \Omega\tilde{\sigma}_{e,h}  S_\delta(E_{e}-E_{h}-\hbar\Omega)\Phi\:,
\end{equation}
at a  given photon frequency $\Omega$,
where $\Phi$ is the photon flux density and
\begin{equation}\label{sigma_tilde}\
 \tilde{\sigma}_{e,h}=\frac{8\pi^2\mathcal F^2}{3 n_{\rm out}}\:\frac{e_0^2}{\hbar c}
|\langle e|\bm{r}|h\rangle|^{2}\;.
\end{equation}
Here $e_0$ is the electron charge, 
$c$ is the speed of light, 
and $\mathcal{F}= 3 n^2_{\mathrm{out}}/ (n^2_{\mathrm{in}} + 2 n^2_{\mathrm{out}})$ 
is the local field factor,\cite{Delerue_book_2004} 
with $n_{\mathrm{in}}$ and $n_{\mathrm{out}}$ being refractive indices of media 
inside and outside the NC, respectively ($\mathcal{F}=0.42$ for Si NC in SiO$_{2}$).
The squared matrix element $|\langle e|\bm{r}|h\rangle|^{2}$ has been calculated
taking into account only inter-atomic matrix elements of coordinate,
which is reasonable for Si.\cite{Cruz1999}
The summation in Eq.~(\ref{Wabs}) is performed over all initial hole states
$h$ and final electron states $e$, for which the transition energy
$E_{e}-E_{h}$ is restricted by the half width $\delta$ of the
normalized Lorentzian $S_\delta(x)$, which appears due to the energy conservation law.
Phenomenological parameter $\delta$ determines the average broadening of the
energy levels of excited states, and it is finite due to interaction with phonons. 
In our calculations we assumed $\delta=5$~meV.

The absorption cross section can be found from Eq.~(\ref{Wabs}):
\begin{equation}\label{sigma_tilde}\
 {\sigma}_{\rm abs}(\hbar\Omega)=\sum_{e,h}\hbar\Omega\tilde{\sigma}_{e,h}   S_\delta(E_{e}-E_{h}-\hbar\Omega).
\end{equation}
In Fig.~2 the optical absorption
cross section is presented as function of the photon energy
$\hbar\Omega$  for individual Si NCs of diameters 1.8, 3.0, and 5.0~nm. 
All the curves clearly demonstrate the 
threshold at the direct band gap of bulk silicon at the $\Gamma$-point of $\bm{k}$-space,  
which is equal to 3.32~eV (3.4~eV in the model we used,\cite{Jancu} see the vertical line in Fig.~2).
The inset in Fig.~2 illustrates  the initial energy distribution of electrons for a NC with $D\approx 3$~nm  after absorption of the photon with the
energy equal to 3.5 eV. Each point at the energy $E_e$
is proportional to absorption cross section and corresponds to the sum
$\sum_{h}\tilde{\sigma}_{e,h} S_\delta(E_{e}-E_{h}-\hbar\Omega)$.
One can see that optical absorption mainly creates hot
electrons and, correspondingly, holes of low energy.

\begin{figure}[b!]
\begin{center}
\includegraphics[width=0.4\textwidth]{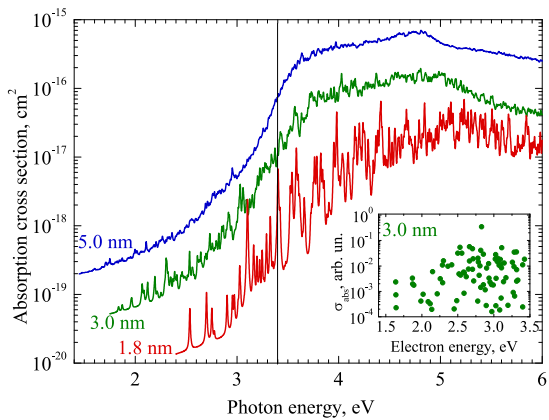}
\end{center}
\caption{(Color online) The optical absorption
cross section  as function of the photon energy $\hbar\Omega$ 
for Si NCs with  diameters 1.8~nm, 3.0~nm, and 5.0~ nm.
The inset illustrates the distribution of electrons
$\sigma_{\rm abs}(E_e)$
in a single NC with $D=3$~nm after absorbing 3.5~eV photon. Vertical line at $3.4$ eV shows the direct band gap energy in bulk Si.
}\label{fig:2}
\end{figure}

Now let us consider the rate of direct spontaneous radiative
recombination of an electron in state $e$ with a hole in state
$h$. It is given by:\cite{Delerue_book_2004}
\begin{equation} \label{rad_time}
  \frac{1}{\tau_{\rm rad}}
		=	\frac{4 e_0^2 |\langle h|\bm{r}|e\rangle|^{2}\mathcal F^2 n_{\rm out}(E_{e}-E_{h})^3 }{3\hbar
  c^3}\;,
\end{equation}
where $E_{e}-E_{h}$ is the transition energy. 
The calculated rate of direct radiative recombination  of hot
electrons with holes situated in four lowest energy levels 
is presented in Fig.~3. The
rate of radiative recombination assisted by emission of optical
phonon is of the order of $10^{4}$~s$^{-1}~$.\cite{Moskalenko_PRB2007} 
Only for such small NC as of 1.8~nm in diameter, 
the rate of direct optical transition from the ground state can reach this value. 
For hot electrons in large NCs ($D=5~$nm, Fig.~3b) fast radiative recombination is
possible only above the direct band gap of bulk Si. 
One can see that most of the fastest transitions  
involve the $\Gamma$-related states~\cite{GammaRelatedStates} of electrons 
(open circles in Fig.~3b), i.e. these transitions are direct.
In smaller NCs fast radiative transitions 
(with rates exceeding $10^{6}$~s$^{-1}$) 
are possible at smaller energies, see Fig.~3a. 
This is explained by stronger quantum confinement and band mixing.
\begin{figure}[t!]
\begin{center}
\includegraphics[width=0.4\textwidth]{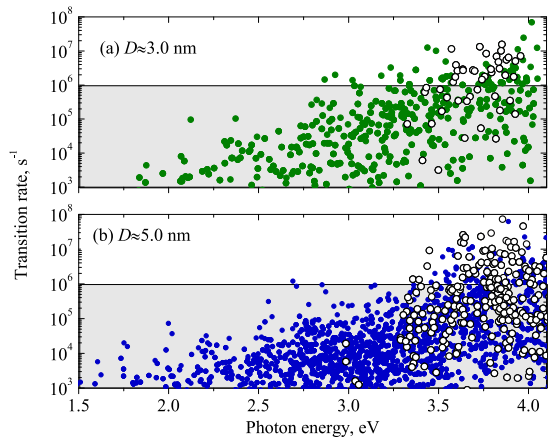}
\end{center}
\caption{(Color online) 
Calculated rates of direct radiative recombination of hot electrons with holes.
Panels (a) and (b) were calculated for NCs with diameters equal to 3.0 and 5.0~nm. 
Open circles correspond to the transitions 
involving $\Gamma$-related electron states.~\cite{GammaRelatedStates}  
Shaded area indicates the relatively slow transitions with 
$\tau_{\rm rad}>10^{-6}$~s.}\label{fig:3}
\end{figure}

Now we turn to the discussion of Auger-like processes
\cite{efros1995,Wang2003} with energy transfer between electron and
hole via Coulomb interaction. 
These processes are very important in case of optical excitation,
when an electron-hole pair  is generated. 
Transition rate is given by:
\begin{equation}\label{eq:auger}
 \frac1{\tau_{\rm aug}}=\frac{2\pi}{\hbar \delta } |\langle f |H_{\rm C}(\bm r_e,\bm r_h)|i\rangle|^2\:.
\end{equation}
Here $\langle f |H_{\rm C}(\bm r_e,\bm r_h)|i\rangle$ is the
matrix element of the electron-hole Coulomb interaction in the
NC, which was evaluated neglecting short-ranged interaction and image forces.
The difference between the transition energies of
electrons  and holes must be smaller than the broadening $\delta $.
We considered Auger-like processes  with initial state of the holes being the ground one, so that the
energy is transferred from hot electrons to the holes.
This is supported by our analysis of initial distribution of hot  carriers after photon absorption, see inset to Fig.~2.
Our calculations indicate very high rates of Auger-like processes for all the electron levels
(except for low energy states in smaller NCs), 
with typical transition times $\tau\lesssim
10^{-13}$~s. 
However, after a hot hole appears in the
NC and the electron is cooled down, the inverse process, with energy
transfer back from the hole to the electron, becomes
possible. 
Such forth and back energy exchange between electrons and holes 
establishes the joint two-particle distribution function. 
While such process does not involve phonons, 
it is elastic and does not lead to energy relaxation. 

\begin{figure}[t!]
\begin{center}
\includegraphics[width=0.45\textwidth]{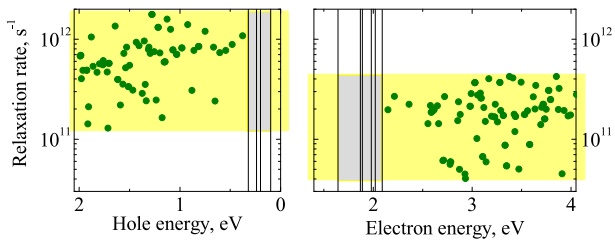}
\end{center}
\caption{Rates of single optical phonon emission by electrons and holes in Si NC ($D=3$~nm). 
Vertical lines illustrate energies of confined carriers in the bottleneck regions, 
where single-phonon relaxation is impossible.}\label{fig:4}
\end{figure}

The energy relaxation takes place only due to phonon-assisted processes. 
The energy gaps between the highly excited levels of size quantization 
are less than the energy of optical phonons and single-phonon relaxation is possible. 
We have considered the energy relaxation of hot electrons and holes 
via emission of single optical phonon. 
The matrix elements of deformation potentials $\langle f |D_{\rm opt}|i\rangle$ have been found in
the tight binding approach.\cite{Jancu} 
Bulk parameters of phonon modes in Si 
have been used with optical phonon energy $\hbar\omega_{\rm opt}=60$~meV. 
Note, that it has been shown in Refs.~\onlinecite{Valentin2008} that for NCs
with diameters $D\gtrsim 2$~nm the interatomic forces remain close
to those in the bulk and there is no dramatic difference in the corresponding phonon spectra.
Energy conservation law is fulfilled due to the dispersion of optical phonons, 
which magnitude was estimated as $\Delta=2\delta\sim 10~$meV.  
Thus, the transition rate is given by 
$1/\tau=(2\pi \hbar/\Delta) |\langle f |D_{\rm opt}|i\rangle|^2$,
if the difference between  the energies of initial and final
states $|E_f+\hbar\omega_{\rm opt}-E_i|$ is smaller than $\delta$, 
and zero otherwise. 
The obtained rates are of the order of $10^{10}\div 10^{11}$~s$^{-1}$ for $D=4~$nm 
and increase up to $10^{12}$~s$^{-1}$ for smaller NCs with $D=2.5~$nm. 
The corresponding rates for holes are slightly higher than for electrons.
The requirement for energy conservation 
forbids relaxation for a considerable number of states,  including the 
highly excited ones,
especially in smaller NCs. 
For instance, for $D=3~$nm such process is allowed only for three quarters of the electron states, 
the calculated rates are presented in Fig.~4. 
Thus, the relaxation of the rest states requires participation of acoustic phonons.
Our preliminary analysis indicates that such processes can be characterized by significantly larger transition times than optical phonon emission.

In conclusion, we have analyzed various radiative and nonradiative processes in Si NCs. 
The results are summarized below.
(i) Absorption of high-energy photons mainly creates hot electrons and cold holes. 
(ii) Auger-like energy exchange processes between electrons and holes, 
taking place due to the Coulomb interaction, are the fastest in the system.
But they do not directly lead to the energy relaxation. 
(iii) Energy relaxation is governed by optical phonon emission 
with the rates of the order of $10^{10}\div 10^{12}$~s$^{-1}$. 
For a considerable number of states in small NCs single-phonon processes 
are ruled out by energy conservation law 
and slower multiphonon relaxation involving acoustic phonons takes place. 
Auger-like energy exchange mediates selection of the fastest available relaxation channels.
(iv) In larger NCs ($D\gtrsim 5~$nm) fast radiative recombination (with rates exceeding $10^{6}$~s$^{-1}$) is
possible only for electrons with the energies close to $3.4$~eV, the direct band gap of bulk silicon.
For smaller NCs fast radiative transitions appear also at smaller energies.

We thank  T.~Gregorkiewicz for initiation of this work and
 C.~Delerue, E.\,L. Ivchenko, A.\,S.~Moskalenko and M.\,O. Nestoklon for very useful discussions. The support by RFBR, programs of RAS, and ``Dynasty'' foundation-ICFPM is gratefully acknowledged.


\begin{thebibliography}{99}

\bibitem{Pavesi}
L.~Pavesi, L.~Dal Negro, C.~Mazzoleni, G.~Franzo, F.~Priolo,
Nature \textbf{408}, 440 (2000).

\bibitem{Kenyon}
A.~J.~Kenyon,
Prog. Quantum Electron. \textbf{26}, 225 (2002).

\bibitem{Katya}
K.~Dohnalov\'{a},  L. Ondi{\v c},  K.~K\r{u}sov\'{a}, I. {Pelant}, J.L.~{Rehspringer}, and R.-R. {Mafouana}, 
J. Appl. Phys. {\bf 107}, 053102 (2010).

\bibitem{JETF}
A.A. Prokofiev, A. S. Moskalenko, I. N. Yassievich, W. D. A. M. de Boer, D. Timmerman,
 H. Zhang, W. J. Buma, T. Gregorkiewicz, JETP Lett. {\bf 90}, 856 (2009).

\bibitem{Nature}
D.~Timmerman, I.~Izeddin, P.~Stallinga, I.~N.~Yassievich, T.~Gregorkiewicz,
Nature Photonics \textbf{2}, 105 (2008).

\bibitem{Jancu}
J.-M. Jancu, R. Scholz, F. Beltram, F. Bassani, Phys. Rev. B,  {\bf 57}, 6493 (1998)

\bibitem{Delerue_2000}
Y.M.Niquet, C.Delerue, G.Allan, M.Lannoo, Phys.Rev. B  {\bf 62}, 5109 (2000)



\bibitem{Delerue_book_2004}{C. Delerue}, M. Lanoo, 
C. Delerue, M. Lanoo, {\it Nanostructures. Theory and Modelling}, (Springer, 2004)

\bibitem{Trani2005}
F. Trani and G. Cantele and D. Ninno and G. Iadonisi,
Phys. Rev. B {\bf 72}, 075423 (2005).

\bibitem{Kovalev2010}
B. Goller and S. Polisski and H. Wiggers and D. Kovalev,
Appl. Phys. Lett. {\bf 97}, 041110 (2010).

\bibitem{Hill}
N. A. Hill,  and K.B. Whaley, J. Electronic Materials, {\bf 25}, 269 (1996).

\bibitem{Moskalenko_PRB2007}
A.S.~Moskalenko, J.~Berakdar,  A.A.~Prokofiev, I.N.~Yassievich,
Phys. Rev.~B \textbf{76}, 085427 (2007).



\bibitem{Cruz1999}
M. Cruz and M. R. Beltr{\'a}n and C. Wang and J. Tag{\"u}e{\~n}a-Mart{\'i}nez, Phys. Rev. B 59, 15381(1999).
\bibitem{efros1995}
Al.L. Efros, V.A.~Kharchenko, and M.~Rosen, Solid State Comm. {\bf 93}, 281 (1995).
\bibitem {Wang2003}
L.-W. Wang, M. Califano, A. Zunger, A. Franceschetti, Phys. Rev. Lett., {\bf 91}, 056404(2003).

\bibitem{Valentin2008}
A. Valentin, J.See, S. Galdin-Retailleau, and P.Dollfus, J.Phys.: Cond. Matter {\bf20}, 145213 (2008)

\bibitem{GammaRelatedStates}
By writing ``$\Gamma$-related'' we refer to those states of confined electrons,
for which the Fourier image of the wave function 
is concentrated near the $\Gamma$ point of the reciprocal space.

\end{thebibliography}
\end{document}